# INTEGRATION OF LATEX FORMULA IN COMPUTER-BASED TEST APPLICATION FOR ACADEMIC PURPOSES

Ikechukwu E. Onyenwe[1], Ebele Onyedinma[1], Onyedika O. Ikechukwu-Onyenwe[1], Obinna Agbata [1], and Faustinah N. Tubo [1]

[1]Department of Computer Science, Nnamdi Azikiwe University, Awka, Nigeria
ie.onyenwe@unizik.edu.ng, eg.osita@unizik.edu.ng, ou.agbata@unizik.edu.ng, ft.tubo@stu.unizik.edu.ng.

## ABSTRACT

*LaTeX is a free document preparation system that handles the typesetting of mathematical expressions smoothly and elegantly. It has become the standard format for creating and publishing research articles in mathematics and many scientific fields. Computer-based testing (CBT) has become widespread in recent years. Most establishments now use it to deliver assessments as an alternative to using the pen-paper method. To deliver an assessment, the examiner would first add a new exam or edit an existing exam using a CBT editor. Thus, the implementation of CBT should comprise both support for setting and administering questions. Existing CBT applications used in the academic space lacks the capacity to handle advanced formulas, programming codes, and tables, thereby resorting to converting them into images which takes a lot of time and storage space. In this paper, we discuss how we solvde this problem by integrating latex technology into our CBT applications. This enables seamless manipulation and accurate rendering of tables, programming codes, and equations to increase readability and clarity on both the setting and administering of questions platforms. Furthermore, this implementation has reduced drastically the sizes of system resources allocated to converting tables, codes, and equations to images. Those in mathematics, statistics, computer science, engineering, chemistry, etc. will find this application useful.*

## KEYWORDS

*LaTeX, CBT, Mathematical formulas, Programming codes, Tables.*

## 1. INTRODUCTION

The use of computers to deliver and mark school assessment is likely to grow in the future: In 2004, a speech by the head of the Qualifications and Curriculum Authority for England (QCA) proposed that, by 2009, "all new qualifications" and "most GCSEs, AS and A2 examinations should be available optionally on-screen" [1]. In recent times, computer-based testing (CBT) has become widely used as an alternative to using the pen-paper method for conducting general knowledge assessments on a subject matter. It was used in a few stages of assessments, such as in education for semester/term quizzes, but as the requirements increased drastically in the post-COVID-19 era including competition, professional, and employment exams; which are used to analyse the knowledge and skills of candidates. For instance, CBT promises a fast, cost-effective, and objective way of sampling students' performance without overburdening (or even involving) teachers. In Examination boards, it seeks to minimize costs, and has the potential to eliminate the cost of printing papers and securely transporting them between schools and markers, and as well remove the need for manual marking. Moreso, it offers the hope of improving the nature and quality of assessment by increasing the range of task types that can be set in high stakes. Apart from the above highlightde hallmarks of CBT, and the provision of platform for conducting mass examinations, it also streamlines the examination process and certification conducted in multiple ways. However, there are challenges as associated with the use of CBT in assessing formular, image and coding prevalent in mathematics, computer science, physical sciences and engineering. In most establishments, computer-based tests in aforementioned areas require a good visual representation of formulas/equations and codes to increase questions readability and clarity. Unfortunately, it is not always the case as most of the questions found in these fields are first typed in a different software and then converted into images in the CBT application. This is a non-trivial task as it takes time and uses a lot of computer memory.

LaTeX was initially invented as a typesetting language for mathematical notation. It is text based and nongraphical in nature. By typing standard text on a keyboard, one can represent all of the mathematical symbols, from the most elementary to the most advanced using a latex editor. A latex editor is a program for creating and editing documents written in the LaTeX markup language. Latex is the most popular tool for writing research papers, books, and other academic or technical documents. It has a variety of features that are optimized for the creation of scientific and technical publications. It enables you to input tables, pictures, and mathematical symbols, among other things. For most researchers, the most useful part of LaTeX is the ability to typeset complex mathematical formulas. LaTeX separates the tasks of typesetting mathematics and typesetting normal text by the use of two operating modes, paragraph and math mode. It allows a few ways of entering math mode. The most common is *$....$*, where the text within the dollar signs is in the math mode environment. Math mode is denoted by using the *\begin{equation} and \end{equation}* commands [7].

The major aspect of this paper is integrating the mathematical formulas of LaTeX to the CBT systems. This enhances the typesetting, editing and rendering of formulas in text-based modes instead of images that is commonly obtainable in most of CBT applications. Thus, proffers solution that better reflects true editor for setting multiple choice questions (MCQ) and answers. This research benefits individuals and organizations that use Mathematics, Physics, Chemistry and Engineering MCQ for assessments.

**2. REVIEW OF RELATED LITERATURE**

This section reviewed previous works related to Latex and formulas for educational purposes, especially in examination.

[2] highlighted that the World Class Test was a QCA/DfES-funded program to identify and engage potentially gifted and talented students, particularly those whose ability was not apparent from their performance on day-to-day classroom activities. The author focussed on two related subject areas - "Mathematics" and "Problem-solving in

mathematics, science, and technology". This work examined some issues that emanated from the use of computer-based assessment of Mathematics in primary and secondary education. As a matter of fact, it considered the usefulness of computer-based assessment for testing "process skills" and "problem-solving". This is discussed through a case study of the World Class Tests project which set out to test problem-solving skills. The study also considered how on-screen "eAssessment" differs from conventional paper tests and how transferring established assessment tasks to the new media might change their difficulty, or even alter what they assess. Writing formulas in LaTeX can be difficult, especially for complex formulas, thus [3] developed MathDeck Formula Editor, which is an interactive formula entry combining LaTeX, Structure Editing, and Search. The research aims to simplify LaTeX formula entry by: 1) allowing rendered formulas to be edited directly alongside their associated LaTeX strings, 2) helping build formulas from smaller ones, and 3) providing searchable formula cards with associated names and descriptions. Equation Editor by [4] allows students to construct equations with Greek letters, exponents, and mathematical symbols. In [5], the authors demonstrated the Formulator project of MathML editor which provides a user with a WYSIWYG editing style while authoring MathML documents with Content or mixed markup. Furthermore, they presented an approach taken to enhance the availability of the MathML editor to end-users, demonstrating an online version of the editor that runs inside a Web browser. According to [6], a formula editor is a computer program that is used to typeset mathematical formulas and mathematical expressions. Typical features of this editor include the ability to nest fractions, radicals, superscripts, subscripts, overscripts, and underscripts together with special characters such as mathematical symbols, arrows, and scalable parentheses. Some systems are capable of re-formatting formulae into simpler forms or adjusting line-breaking automatically while preserving the mathematical meaning of a formula. Considering [8], the author describes an approach called eqExam which is suitable and can only be effective with LaTeX packages, to create an online/interactive exam that supports PDF. The eqExam package provides the usual features for creating tests, quizzes, and homework. With it, you can create multiple-choice, True/False, short fill-in-the-blank, and extended response questions. [9] Introduces the functionalities of RndTexExams, which was designed to minimize visual cheating and facilitate the grading of printed exams. With RndTexExam, it is easier to create exams with randomized spatial content by automatically changing the order of questions, their textual content, and the order of answers. It also includes a specialized function that, when combined with a cloud-based service, makes it easy to grade a large number of multiple-choice exams built with RndTexExams. The package can be used to statistically test for cheating based on student answer sheets. [10] uses the Place Value Assessment Tool (PVAT) and its online equivalent, the PVAT-O. In the PVAT-O creation, Multiple technologies including HyperText Markup Language (HTML5), Javascript, and PHP: Hypertext Preprocessor (PHP) were used to create the PVAT-O assessment. The mathematical content and format of each PVAT-O item were as close as possible to the equivalent PVAT items. However, some items required the inclusion of computer-based features. For example, a 'drag and drop' feature was used in items requiring students to place numbers in order from smallest to largest, and 'radio buttons' were used in multiple-choice items. [11] Introduces Numbas; a new SCORM-2004 compliant as an open-source and multi-platform e-assessment and e-learning system. They focused on rich formative e-assessment and learning; blending powerful mathematical and statistical functionality with its unique browser and client-based design, bringing into play the full capability and resources of the internet. Numbas is entirely written in Javascript, along with some small Python scripts which compile examination packages for distribution. They used MathJax to solve the problem of displaying mathematical notation on the web. Authors use a very simple structured data format, similar to JSON, to create exams. All content displayed to the user is written as simple HTML or Textile, with LaTeX used for mathematical notation. Formative assessment benefits both students and teaching academics [12]. Most especially, formative assessment in mathematics subjects enables both students and teaching academics to assess individual performance and understanding through students' responses. Over the years, educational technologies and learning management systems (LMSs) have been used to support formative assessment design. In mathematics, this is problematic because of the inflexibility of LMS and educational technology tools. Hence, they proposed a new method of creating mathematics formative assessments using LaTeX and PDF forms in conjunction with a computer algebra system (e.g., Maple), independent of an LMS. The method generates individualized assessments that are automatically marked. Results show that the method provides the teaching academic with a more efficient way of designing formative mathematics assessments without compromising the effectiveness of the assessment task.

## 3. EXPERIMENTAL METHODOLOGY

This paper on integrating LaTeX technology into CBT applications to enable seamless manipulation and accurate rendering of tables, programming codes, and equations to increase readability and clarity on both setting and administering of questions platforms.

Integrating LaTeX into Computer-Based Test allows for mathematical expressions, symbols, rich text formatting, and code input. It enables precise and professional rendering of mathematical symbols and equations, ensuring accurate representation of mathematical concepts in assessments. It also allows creation of complex mathematical structures and symbols, providing flexibility for test designers to include a wide range of mathematical problems and expressions. Its scalability makes it suitable for a wide range of educational levels and subjects that involves mathematical notations. In order to achieve this integration, we implement a LaTeX compiler into the CBT system to interpret and render LaTeX code in HTML representations. Moreso, we develop mechanisms for accepting LaTeX input from CBT creators and rendering LaTeX-formatted content for CBT test takers. This includes designing input interface for test creation using CBT editor, output interfaces for CBT test presentation using manager and client server and implementing storage system for storing LaTeX-formatted test content using the backend server and retrieval mechanisms to efficiently access and present this content during assessments.

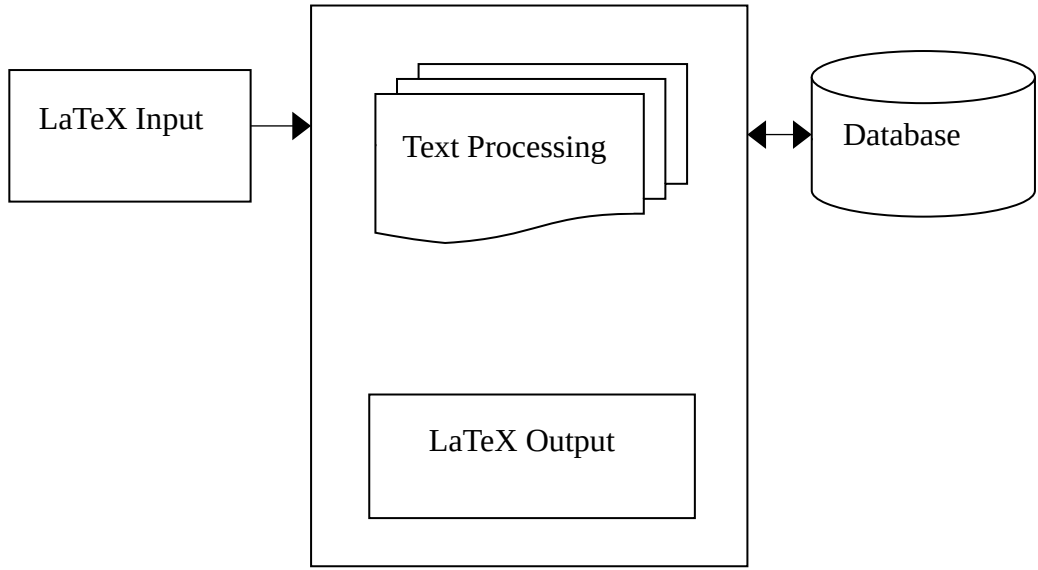

Figure 1. Framework of the CBT application

Figure 1 shows the general architecture of how CBT tests are created using the experimental tools explained in subsection 3.1.

### 3.1. Experimental Tools and Setups

The following experimental tools were used in the integration of LaTeX formulas into a computer-based test application such as Django and associated Python libraries, LaTeX, vue.js.

**Django** is a high-level web framework for building web applications using the Python programming language. It follows the model-view-controller (MVC) architectural pattern and encourages the use of reusable code by providing a clean and pragmatic design. Django includes an Object-Relational Mapping (ORM) system for database access, a templating engine for rendering views, and a built-in administration panel, among other features. Django officially supports databases such as PostgreSQL, MariaDB, MySQL, Oracle, and SQLite [13]. For this reseach, we used PostgreSQL.

**LaTeX** is a typesetting system widely used for the production of scientific and mathematical documents due to its exceptional handling of complex equations and formatting. It is not a word processor like Microsoft Word or Google Docs; instead, it is a markup language and document preparation system.

**Vue.js** is an open-source model–view–viewmodel frontend JavaScript library for building user interfaces and single-page applications. It was created by Evan You, and is maintained by him and the rest of the active core team members. Vue supports a library for mathematical equation handling using Latex (latex2js). Using this library, we built an interactive user interfaces that accepts latex code inputs given a formula field. This provides a platform for examiners in the discipline of mathematics, physics, chemistry, computer and engineering disciplines to express mathematical equations and other advanced formulas to their students seamlessly. The presentation provides deep, educational experiences for the students to engage more in the science questions without any form of blurring.

Finally, the system makes use of server-client architecture. There are Manager and students' Exam dashboards to represent the server and client respectively. Creating and setting of exams happens at the Manager's side. Once a student logs in, it shows a dashboard where the student can select the exam he/she is writing if started.

**3.2 Data Collection**

Data collection is a crucial step in the research process. This involves gathering information or data for analysis and interpretation. When integrating LaTeX into computer-based tests, the focus of data collection often revolves around user input and the rendering of LaTeX content. For user input, LaTeX inputs are collected from users when they submit responses or create questions. While for rendering LaTeX content, either client-side solutions (e.g., MathJax) or server-side rendering libraries (e.g., Django-latex) is used. In this research, we used mathematical basic courses from Nnamdi Azikiwe University, Awka, Nigeria such as Math 101, Math 102, Math 103, Feg 303 and so on to test the system.

**4. RESULT DISCUSSION**

This section discusses the results of integrating LaTeX into a computer-based test (CBT) application. Results show that users can easily input and format mathematical expressions, equations, and symbols. Again, the system provides a standardized and efficient way to represent mathematical content, ensuring the overall quality of mathematical notation in the test environment. It significantly improved mathematical representation by offering a robust system for expressing complex mathematical notations, equations, and symbols.

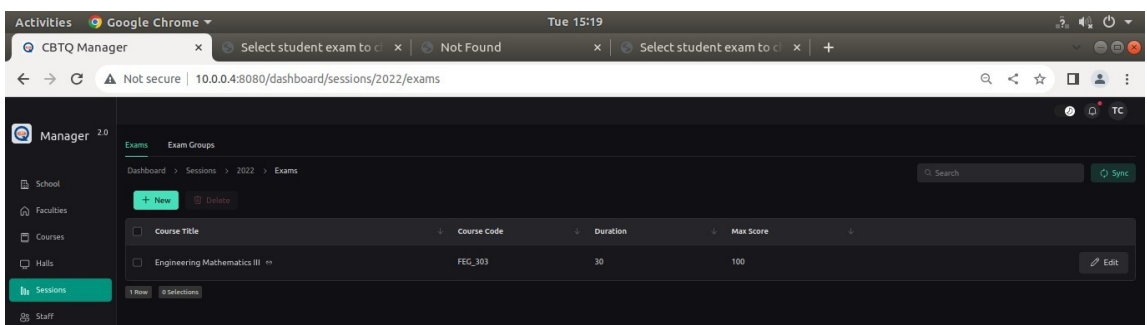
Fig 3.1 Create an exam

Figure 3.1 creates an exam (e.g., Engineering Mathematics III) using the Vue.js developed manager. As shown in Figure 1 and implemented in Figure 3.2, the test creator input, received the LaTeX-formatted content in the input field and it was processed alongside the associated HTML tags using the Latex2JS libraries and Django. Figure 3.2 images are labeled accordingly starting with #1 showing when an examiner inserted the Formula/Latex field developed using Latex2JS Vue library; #2 is Latex symbols/characters the examiner entered; #3 renders the characters entered in #2 and generating a readable texts; #4 shows inserted formula/latex field and formula/latex characters for option A; #5 shows rendered results for option A. All these happened at the server-side of the application called the Manager. Thereafter, the processed LaTeX content is stored in the database. CBT test contents were generated based on the processed LaTeX information, the system generates the actual test content and renders it to the test taker only if the test taker is a valid user. The final test content includes the LaTeX-rendered mathematical expressions which are presented to the user at the client-side of the CBT application (see Figure 4).

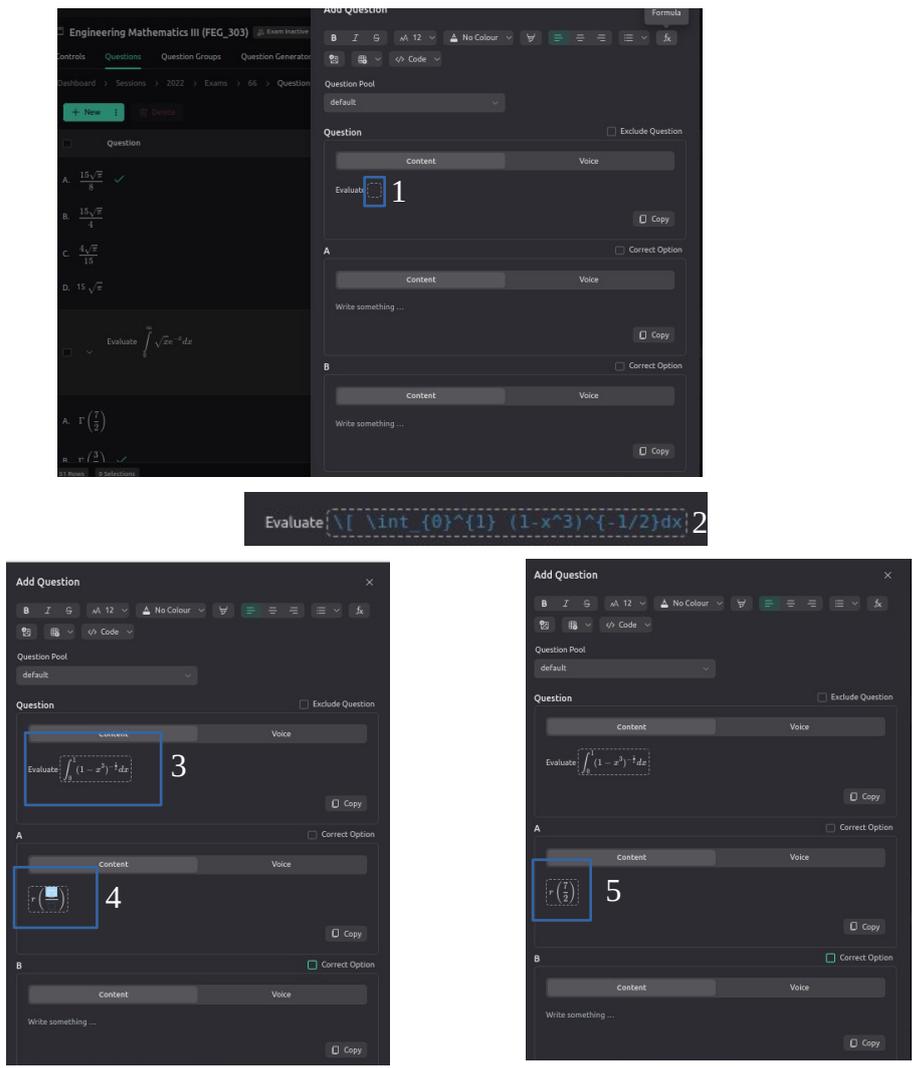

Fig 3.2 Formula/LaTeX field

Figure 3.2 shows the LaTeX editor environment where mathematical equations were entered and the respective options, and of course the correct answer was also selected.

The client-side administers exams to the students. It displays the contents of the database as entered using CBT editor illustrated in Figures 1 and 3.2. The database, as discussed in section 3.1, stores the latex/formula symbols along with the associated HTML tags as scripts (see Table 1), which on request, renders the scripts by generating readable mathematical/chemical expressions as shown in Figure 4.

```
<template>
    <div class="min-h-screen">
        <body>
            <math-field read-only style> \sum_{k=1}^{2}a*b^2 </math-field>
        </body>
    </div>
</template>
```
Table 1: Sample of the scripts stored in the database at the server-side

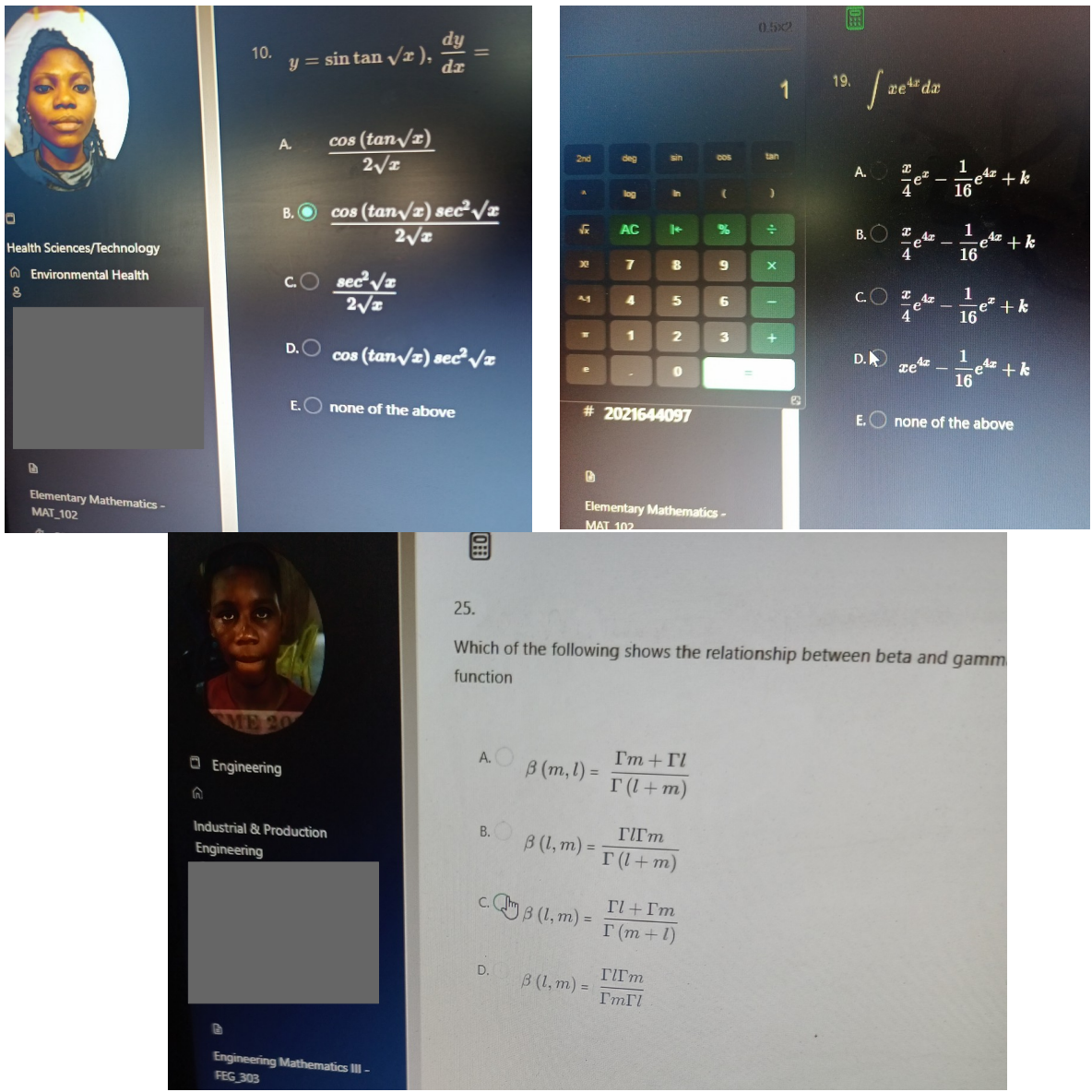

Figure 4: Mathematical expressions generated as a result of the activities of Figure 3.2. These are texts not images as the case of most CBTs where these expressions are rendered as images.

## 5. CONCLUSION

The Integration of LaTeX into Computer-Based Tests marks a significant stride towards enriching the assessment of mathematical and scientific knowledge. By leveraging LaTeX's robust typesetting capabilities, we enhanced the precision and clarity of mathematical expressions, thereby elevating the overall quality of content representation in the digital testing environment. Throughout this exploration, we delved into crucial aspects such as addressing rendering challenges, implementing security measures, and prioritizing user experience. These considerations are pivotal in ensuring a seamless and effective integration that benefits both educators and students. As we conclude, it is evident that embracing LaTeX in computer-based tests not only aligns with the evolving technological landscape but also fosters a more nuanced evaluation of mathematical proficiency. The journey from syntax understanding to real-time rendering feedback, coupled with insights from successful case studies, equips stakeholders with the knowledge to navigate challenges and optimize the integration process. Looking ahead, continued collaboration between educational institutions, test developers, and technology providers will play a pivotal role in refining and advancing the integration of LaTeX into computer-based testing, ultimately contributing to a more sophisticated and comprehensive assessment experience for learners across diverse mathematical discipline.

**Authors**

| | Short Biography | Photo 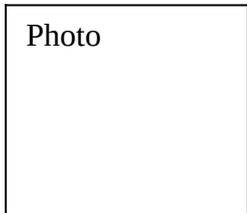 |